\documentclass[10pt,twocolumn,aps,english,pra,superscriptaddress]{revtex4}
\usepackage{amsfonts}
\usepackage{mathrsfs}
\usepackage{amsmath}
\usepackage{amssymb}
\usepackage{amsthm}
\usepackage{babel}
\usepackage{graphicx}
\usepackage{float}
\usepackage{subfigure}
\usepackage[a4paper]{geometry}
\usepackage{caption}

\newcommand{\myref}[1]{Eq.\hspace{0.3em}\eqref{#1}}
\newcommand{\mycite}[1]{Ref.\hspace{0.3em}\cite{#1}}

\newcommand{\myfig}[1]{Fig.\hspace{0.3em}\ref{#1}}
\newcommand{\bgeq}{\begin{equation}}
\newcommand{\edeq}{\end{equation}}

\begin{document}

\title{Entropic uncertainty relations for general SIC-POVMs and MUMs}

\author{Shan Huang}
\affiliation{School of Physics, Nanjing University, Nanjing 210093, China}
\affiliation{Institute for Brain Sciences and Kuang Yaming Honors School, Nanjing University, Nanjing 210023,
China}

\author{Zeng-Bing Chen}
\email{zbchen@nju.edu.cn}
\affiliation{School of Physics, Nanjing University, Nanjing 210093, China}

\author{Shengjun Wu}
\email{sjwu@nju.edu.cn}
\affiliation{School of Physics, Nanjing University, Nanjing 210093, China}
\affiliation{Institute for Brain Sciences and Kuang Yaming Honors School, Nanjing University, Nanjing 210023,
China}

\date{\today}

\begin{abstract}
We construct inequalities between R\'{e}nyi $\alpha$-entropy and the indexes of coincidence of probability distributions, based on which we obtain improved state-dependent entropic uncertainty relations for general symmetric informationally complete positive operator-valued measures (SIC-POVMs) and mutually unbiased measurements (MUMs). We show that our uncertainty relations for general SIC-POVMs and MUMs can be tight for sufficiently mixed states, and, moreover, comparisons to the numerically optimal results are made via information diagrams.
\end{abstract}

\maketitle

\section{INTRODUCTION}
Incompatible observables cannot be measured with certainty simultaneously, though contrary to the general cognition of the physical world based on macroscopic experience, this is a fundamental element of quantum mechanics. Heisenberg made the first statement of this kind of uncertainty of quantum mechanics \cite{heisenberg}, and formulated the first uncertainty relation
\begin{equation}\Delta P \Delta Q\geq\frac{\hbar}{2},\end{equation}
where $\Delta P$ and $\Delta Q$ denote the standard deviation of momentum and position along the same direction respectively. Robertson generalized it to two arbitrary observables \cite{Robertson}
\begin{equation}\Delta X\Delta Y\geq \frac{1}{2}|\langle \psi|[X,Y]|\psi\rangle|,\label{Robertson}\end{equation}
where [X,Y] denotes the commutator between X and Y.

Though clear and elegant enough, the standard deviation way of measuring uncertainty sometimes can be quite strange \cite{Deutsch,Dam,Bia} and it turns out to be inappropriate in applications of information theory.
On the other hand, entropy is found to be a more universal and effective measure of uncertainty \cite{Deutsch,Coles,Friedland,JJ}, and entropic uncertainty relations (EURs) have many applications in quantum information theory. The lower bound on conditional min-entropy can characterize how much randomness one can extract from a source \cite{qrandom}, and more over, as entanglement between two systems reduces the uncertainty (lower bound on entropy) of measurements performed on one system provided that the other one is accessible, EURs are useful in entanglement witnessing \cite{entangle1,entangle2,entangle3}. EURs are also important in the security proof of quantum cryptography as they measure how much information is possibly leaked to an eavesdropper \cite{crypt1,crypt2}. (See more applications in the review \cite{CBTW} and references therein.)

Any projective measurement made in one base cannot reveal any information stored in bases that are mutually unbiased to it, and this property endows mutually unbiased bases (MUBs) with a special role in quantum information theory. Based on the work of Deutsch \cite{Deutsch} and Kraus \cite{Kraus}, Maassen and Uffink proved the famous tight state-independent uncertainty relation for two MUBs in terms of Shannon entropy \cite{MU}, and a generalization to multiple MUBs has also been explored \cite{ID,J3,J5,BW,WW,WYM}. However, an analytic construction of more than three MUBs in general dimensions has not been found, and the existence of complete MUBs in non-prime-power dimensional spaces such as $d=6$ is still an open question \cite{Ben}.

While general symmetric informationally complete positive operator-valued measures (SIC-POVMs) \cite{JKAC} and mutually unbiased measurements (MUMs) \cite{mum} are positive-operator-valued measures with interesting properties similar to MUBs, and a complete set of them can be constructed analytically in all dimensions \cite{gsicexist,mum}, uncertainty relations have been naturally generalized to take into consideration more generalized measurements \cite{Raste,indexofgsic,indexofmum,WWS} like them. In two recent works EURs are also constructed from quantum designs \cite{design1,design2}. In this paper, we focus on uncertainty relations for SIC-POVMs and MUMs and deal with them under a unified framework.

This paper is structured as follows. In Sec. II we introduce some necessary notations and review the concepts of entropy, SIC-POVMs and MUMs. In Sec. III we propose entropic uncertainty relations for general SIC-POVMs, and in Sec. IV uncertainty relations for MUMs are constructed. In Sec V, we make further discussions and draw a brief conclusion.
\vspace{-0.4cm}
\section{Preliminaries}

A positive operator-valued measure (POVM) $\mathcal{P}$ on a $d$-dimensional Hilbert space $\mathcal{H}_d$ consists of a set of positive semi-definite operators that sum up to identity: $\mathcal{P}=\big\{P_i\big|P_i\geq 0,\sum_iP_i={\bf1}_d\big\}$. The probability  distribution induced by performing a POVM measurement $\mathcal{P}$ on a quantum state $\rho$ is denoted by $\mathscr{P}=\big(p_1,p_2,\cdots\big)$, where $p_i=Tr(P_i\rho)$ is the probability of obtaining the $ith$ result; the corresponding index of coincidence is defined as the sum of the squares of the probabilities, i.e.,
\begin{equation}
I(\mathcal{P}|\rho)=I(\mathscr{P})=\sum_ip_{i}^2 .
\end{equation}
\vspace{-1cm}
\begin{table}[H]
\caption{Notations and meaning.}
\begin{tabular}{|c|l|l|}
\hline
1& $\rho$&Density matrix\\\hline
2&$L$&The length of a probability distribution\\\hline
3&$\mathcal{H}_d$&$d$-dimensional Hilbert space\\\hline
4& ${\bf 1}_d$&$d$-dimensional Identity matrix \\\hline
5&{\bf P}&A boldfaced letter is, if not specified \\
&{}&otherwise, a finte set of POVMs\\\hline
6&$\mathscr{P}$&A probability distribution\\\hline
7&$I(\mathscr{P})$&Index of coincidence of $\mathscr{P}$\\\hline
8&$I({\bf P}|\rho$)&The sum of indexes of coincidence  \\
&{}&induced by performing {\bf P} on $\rho$\\\hline
9&$H_{\alpha}({\bf P}|\rho)$&Sum of R\'{e}nyi-$\alpha$ entropies for the  \\
{}&{}&measurements {\bf P} performed on $\rho$\\\hline
{}&{}&Two probability distributions over $L$ \\
10&$\mathscr{P}_{x}^L[c]$,$\mathscr{P}_y^L[c]$ & outcomes defined in \myref{px} and  \\
{}&{}&\myref{py}, and $I(\mathscr{P}_{x}^L[c])=I(\mathscr{P}_{y}^L[c])=c$\\\hline
\end{tabular}\label{table}
\end{table}

The Shannon entropy of $\mathscr{P}$, defined by $H(\mathscr{P})=H(\mathcal{P}|\rho)=-\sum_{i=1}^d p_{i}\log_{2}{p_{i}}$, gives a measure of the uncertainty for the measurement outcomes.
R\'{e}nyi generalized it to a family of entropies \cite{Renyi}
 \begin{align*}
 H_{\alpha}(\mathcal{P}|\rho)=\frac{1}{1-\alpha}\log_2\left(\sum_{i=1}p_i^\alpha\right),\hspace{0.5em}(\alpha>0,\alpha\neq1)
 \end{align*}
which reduces to Shannon entropy in the limitation
$\lim_{\alpha\to1}H_{\alpha}(\mathcal{P}|\rho)=H_1(\mathcal{P}|\rho)=H(\mathcal{P}|\rho)$. Following \mycite{HT}, we call the range of the map $\mathscr{P}\rightarrow \big[I(\mathscr{P}),H_{\alpha}(\mathscr{P})\big]$ \textit{information diagrams}.

For any finite set of POVMs $\textbf{P}=\{\mathcal{P}^1,\mathcal{P}^2,\cdots\}$ performed on $\rho$, we consider the sum of indexes of coincidence
\begin{equation}
I(\textbf{P}|\rho)=\sum_{m=1}^{|\bf{P}|}I(\mathcal{P}^m|\rho),
\end{equation}
and the sum of entropies $H_{\alpha}({\bf P}|\rho)=\sum_mH_{\alpha}(\mathcal{P}^m|\rho)$.

Table I contains some notations that are frequently used in this paper.

\subsection{Symmetric informationally complete POVM}

A POVM on $\mathcal{H}_d$ is said to be symmetric informationally complete (SIC-POVMs) \cite{JKAC} if it consists of $d^2$ rank-1 operators $\textbf{S}=\big\{S_i\big\}$ such that $Tr(S_iS_j)=\frac{d\delta_{ij}+1}{d^2(d+1)}$. From the geometric point of view, with $S_i=\frac{1}{d}|\phi_i\rangle\langle\phi_i|$, SIC-POVM comprises of $d^2$ subnormalized equiangular vectors $\big\{\frac{1}{d}|\phi_i\rangle\big\}$ in $ C^d$ as $|\langle\phi_i|\phi_j\rangle|^2=\frac{d\delta_{ij}+1}{d+1}$ and $\sum_{i=1}^{d^2}\frac{1}{d}|\phi_i\rangle\langle\phi_i|={\bf1}_d$.
Although research is still ongoing to prove or disprove the existence of SIC-POVMs for general $d$, analytic and numerical results confirmed its existence for dimensions up to 67 \cite{sicexis}.

SIC-POVMs is informationally complete, as when performed on a system the resulting probability distributions fully reveal all the information of the corresponding density matrix. More concretely, any density matrix $\rho$ can be constructed from the probabilities $\{p_j\}$ induced by SIC-POVM, and with $Tr(\rho S_j)=p_j$ there is $\rho=\sum_jp_j\left[d(d+1)S_j-{\bf 1}_d)\right]$ \cite{indexofgsic}.\\

By generalizing the method proposed in \mycite{WYM}, Rastegin obtained  \cite{Raste}
\begin{align}
I(\textbf{S}|\rho)=\sum_{i=1}^{d^2}p_i^2=\frac{1+Tr(\rho^2)}{d(d+1)}\label{indexsic},
\end{align}
where $p_i=Tr(\rho S_i)$.

Generalizations of SIC-POVM to that with elements of any rank have been explored in Refs. \cite{gsic1,gsic2}, and in \mycite{gsicexist} the authors proved the existence of general SIC-POVMs in all dimensions by giving the explicit construction. Any general SIC-POVM $\textbf{S}_g=\big\{S_i\big\}$ $(i=1,2\cdots, d^2)$ is a POVM satisfying
\begin{align*}
&Tr(S_iS_i)=a,\left(\forall i,1/d^3<a\leq 1/d^2\right)\\
& Tr(S_iS_j)=\frac{1-ad}{d(d^2-1)}\hspace{0.5em}(\forall i\neq j).
\end{align*}

It is shown in \mycite{indexofgsic} that
\begin{align}
I(\textbf{S}_g|\rho)=\frac{(ad^3-1)Tr(\rho^2)+d(1-ad)}{d(d^2-1)}\label{indexgsic}
\end{align}

\subsection{Mutually unbiased measurements}

We say two orthonormal bases $\{|b_i^1\rangle\}$ and $\{|b_j^2\rangle\}$ $(1\leq i,j\leq d)$ in $\mathcal{H}_d$ are mutually unbiased bases (MUBs) \cite{I,WF,KR,PR} if the inner products between their basis vectors satisfy $|\langle b^{1}_i|b^{2}_j\rangle|=\frac{1}{\sqrt{d}}$ $(\forall 1\leq i,j\leq d)$. For any $d\geq2$, one can find at least three MUBs and at most $d+1$ MUBs (an informationally complete set of MUBs). A complete set of MUBs can always be found if $d$ is the power of a prime number, while it is still an open question what's the maximal number of MUBs in general \cite{Ben}.

According to \cite{WYM}, for a set \textbf{B} of MUBs in $\mathcal{H}_d$,
\bgeq
I(\textbf{B}|\rho)\leq Tr(\rho^2)+\frac{|\textbf{B}|-1}{d}.\label{indexmub}
\edeq
Introduced as generalizations of MUBs, mutually unbiased measurements (MUMs) \cite{mum}  are a set of POVMs $\textbf{P}=\{\mathcal{P}^1,\mathcal{P}^2,\cdots\}$ with each $\mathcal{P}^m$ containing $d$ elements $\mathcal{P}^m=\big\{P^m_1,\cdots,P^m_d\big\}$ and satisfy
 \begin{align*}
Tr\big(P^m_i\big)=1,\hspace{1em}&Tr\big(P^m_iP^{m'}_j\big)=\kappa\delta_{ij}\delta_{mm'}+\\
&(1-\delta_{ij})\delta_{mm'}\frac{1-\kappa}{d-1}+(1-\delta_{mm'})\frac{1}{d},
\end{align*}
where $\kappa$ $(\frac{1}{d}<\kappa\leq 1)$ is called the efficiency parameter. Note that the case $\kappa=1$ corresponds with projective measurements consisting of mutually unbiased bases.

For any set \textbf{P} of MUMs on $\mathcal{H}_d$ there is \cite{indexofmum,mum}
\bgeq
I(\textbf{P}|\rho)\leq \frac{|\textbf{P}|}{d}+\frac{\kappa d-1}{d(d-1)}\left[dTr(\rho^2)-1\right],\label{indexmum}
\edeq
and if \textbf{P} is complete,
\begin{align}
I(\textbf{P}|\rho)=\frac{d+1}{d}+\frac{\kappa d-1}{d(d-1)}\left[dTr(\rho^2)-1\right].\label{indexcmum}
\end{align}

\section{Uncertainty relations for general SIC-POVMs}

In the following discussions we always arrange the probabilities in a probability distribution in descending order and ignore the probabilities being zero as they do not contribute to entropy, and we will frequently consider the two kinds of distributions described below. For any integer $L\geq2$ and $\forall c\in[\frac{1}{L},1]$, $\mathscr{P}_x^L[c]$ and $\mathscr{P}_y^L[c]$ are two  probability distributions over $L$ outcomes, the indexes of coincidence of which are both $c$:
\begin{widetext}
\begin{gather}
\mathscr{P}_x^L[c]=\Big(\frac{1+\sqrt{(Lc-1)(L-1)}}{L},(L-1)\odot\frac{1-\sqrt{(Lc-1)/(L-1)}}{L}\Big),
\hspace{6em}I(\mathscr{P}_x^L[c])=c;\label{px}\\
\mathscr{P}_y^L[c]=\Big((\lceil1/c\rceil-1)\odot\frac{1+\sqrt{(\lceil\frac{1}{c}\rceil c-1)/(\lceil\frac{1}{c}\rceil-1)}}{\lceil\frac{1}{c}\rceil},\frac{1-\sqrt{(\lceil\frac{1}{c}\rceil c-1)(\lceil\frac{1}{c}\rceil-1)}}{\lceil\frac{1}{c}\rceil}\Big),\hspace{1em}I(\mathscr{P}_y^L[c])=c\label{py}
\end{gather}
\end{widetext}
where $\lceil\frac{1}{c}\rceil$ is the smallest integer that $\geq\frac{1}{c}$ and $l\odot p$ is shorthand for $l$ probabilities being $p$. Note here the number of nonzero probabilities in $\mathscr{P}_y^L[c]$ is $L,L-1,\cdots,$ respectively when $c\in[\frac{1}{L},\frac{1}{L-1})$, $[\frac{1}{L-1},\frac{1}{L-2}),\cdots,$ i.e., $\lceil1/c\rceil$. Two examples of distributions over four outcomes are presented in \myfig{figdis}

We show in Appendix A the following theorem.

\emph{Theorem 1.} For any discrete probability distribution $\mathscr{P}$ over $L$ outcomes there is
$H_\alpha\big(\mathscr{P}_y^L[I(\mathscr{P})]\big)\leq H_\alpha\big(\mathscr{P}\big)\leq H_\alpha\big(\mathscr{P}_x^L[I(\mathscr{P})]\big)$ for $\alpha\in(0,2]$, and $H_\alpha\big(\mathscr{P}_x^L[I(\mathscr{P})]\big)\leq H_\alpha\big(\mathscr{P}\big)\leq H_\alpha\big(\mathscr{P}_y^L[I(\mathscr{P})]\big)$
 for $\alpha\in[2,+\infty)$, where $H_\alpha\big(\mathscr{P}\big)$ is the R\'{e}nyi-$\alpha$ entropy of $\mathscr{P}$ and $I(\mathscr{P})$ is the index of coincidence of $\mathscr{P}$.

 Thus $H_\alpha\big(\mathscr{P}_x^L[c]\big)$ and $H_\alpha\big(\mathscr{P}_y^L[c]\big)$ are boundary curves of the diagram of $I(\mathscr{P})\text{-}H_\alpha(\mathscr{P})$. The case $L=4$ is shown in \myfig{rangeofentropy} as an example.

The gray (thick) solid line is the graph of $H_2(\mathscr{P})=-\log_2[I(\mathscr{P})]$. The upper bound (UB) on Shannon entropy (blue dashed line) and the lower bound (LB) on R\'{e}nyi 5-entropy (orange dashed dotted line) are respectively given by $H(\mathscr{P}_x^4[c])$ and $H_5(\mathscr{P}_x^4[c])$. At the same time, the lower bound on Shannon entropy (blue solid line) and the upper bound on R\'{e}nyi 5-entropy (red dotted line) are respectively given by $H(\mathscr{P}_y^4[c])$ and $H_5(\mathscr{P}_y^4[c])$.

We should emphasize that Theorem 1 is a generalization of the Shannon entropic bounds obtained earlier in Refs. \cite{J3,HT} to R\'{e}nyi entropy. With Theorem 1 we immediately have the R\'{e}nyi $\alpha$-entropy $H_\alpha(\textbf{S}_g|\rho)$ for performing any general SIC-POVMs with parameter $a$ on $\mathcal{H}_d$ would satisfy

\begin{figure}[H]
\includegraphics[width=0.48\textwidth]{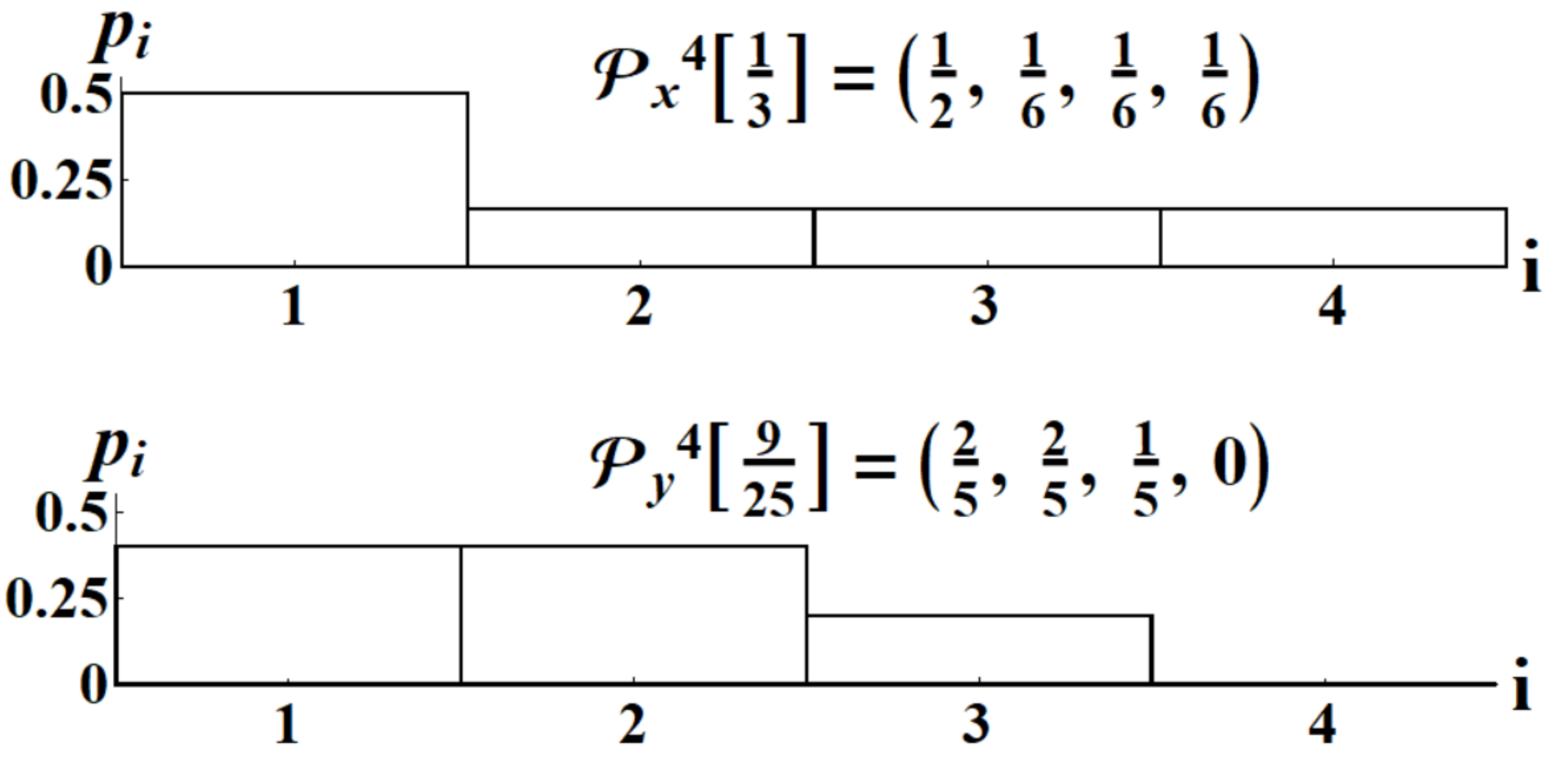}
\caption{Illustrations of distributions over four outcomes} \label{figdis}
\end{figure}
\vspace{-2em}
\bgeq
(2-\alpha)H_\alpha(\textbf{S}_g|\rho)\leq (2-\alpha)H_\alpha(\mathscr{P}_x^{d^2}\left[I(\textbf{S}_g|\rho)\right]),\label{urforgsicpovm1}
\edeq
\vspace{-1.8em}
\bgeq
(2-\alpha)H_\alpha(\textbf{S}_g|\rho)\geq (2-\alpha)H_\alpha(\mathscr{P}_y^{d^2}\left[I(\textbf{S}_g|\rho)\right]),\label{urforgsicpovm2}
\edeq
where $IC(\textbf{S}_g|\rho)$ is given by \eqref{indexgsic}. This is the best result that can be obtained based on \eqref{indexgsic} only, hence uncertainty relations constructed from \eqref{indexgsic} such as those proposed in \mycite{Raste,indexofgsic} cannot be stronger than our results. In the case $\alpha\to\infty$, \myref{urforgsicpovm1} reduces to the result proposed previously by Rastegin \cite{indexofgsic},
\begin{align}
&H_\infty(\textbf{S}_g|\rho)\nonumber\\
\geq&2\log_2d-\log_2\Big[1+\sqrt{ad^3-1}\sqrt{Tr(\rho^2)d-1}\Big].
\end{align}

Now we show \eqref{urforgsicpovm1} and \eqref{urforgsicpovm2} are tight respectively when $Tr(\rho^2)\in[\frac{1}{d},d^2 a]$ and $Tr(\rho^2)\in[\frac{1}{d},\frac{d-2+ad^2}{(d-1)^2}]$. We only need to show the probability distributions $\mathscr{P}_{y/x}^{d^2}\left[I(\textbf{S}_g|\rho)\right]$ can be achieved by some positive semi definite matrix in the form $\rho=\sum_ix_iS_i$, where $x_i=\frac{d(da-1)+d(d^2-1)p_i}{d^3a-1}$ is the solution to $Tr(\rho)=1$ and $Tr(\rho S_i)=p_i\in\mathscr{P}_{y/x}^{d^2}\left[I(\textbf{S}_g|\rho)\right]$. For \eqref{urforgsicpovm1}, when $Tr(\rho^2)\leq d^2a$ we have $x_1\geq x_2=\cdots=x_{d^2}\geq0$, obviously $\rho\geq0$. As for \eqref{urforgsicpovm2}, we have $x_1=\cdots=x_{d^2-1}\geq x_{d^2}$, as $\sum_iS_i={\bf1}_d$, then $\forall |\phi\rangle\in\mathcal{H}_d$, $\langle\phi|\rho|\phi\rangle=\langle\phi|\sum_ix_iS_i|\phi\rangle=x_1+(x_{d^2}-x_1)Tr(S_{d^2}|\phi\rangle\langle\phi|)\geq x_1+(x_{d^2}-x_1)/d\geq0$, thus $\rho$ is a density matrix.

By random sampling over density matrices on $\mathcal{H}_3$,
\vspace{-0.8em}
\begin{figure}[H]
\includegraphics[width=0.48\textwidth]{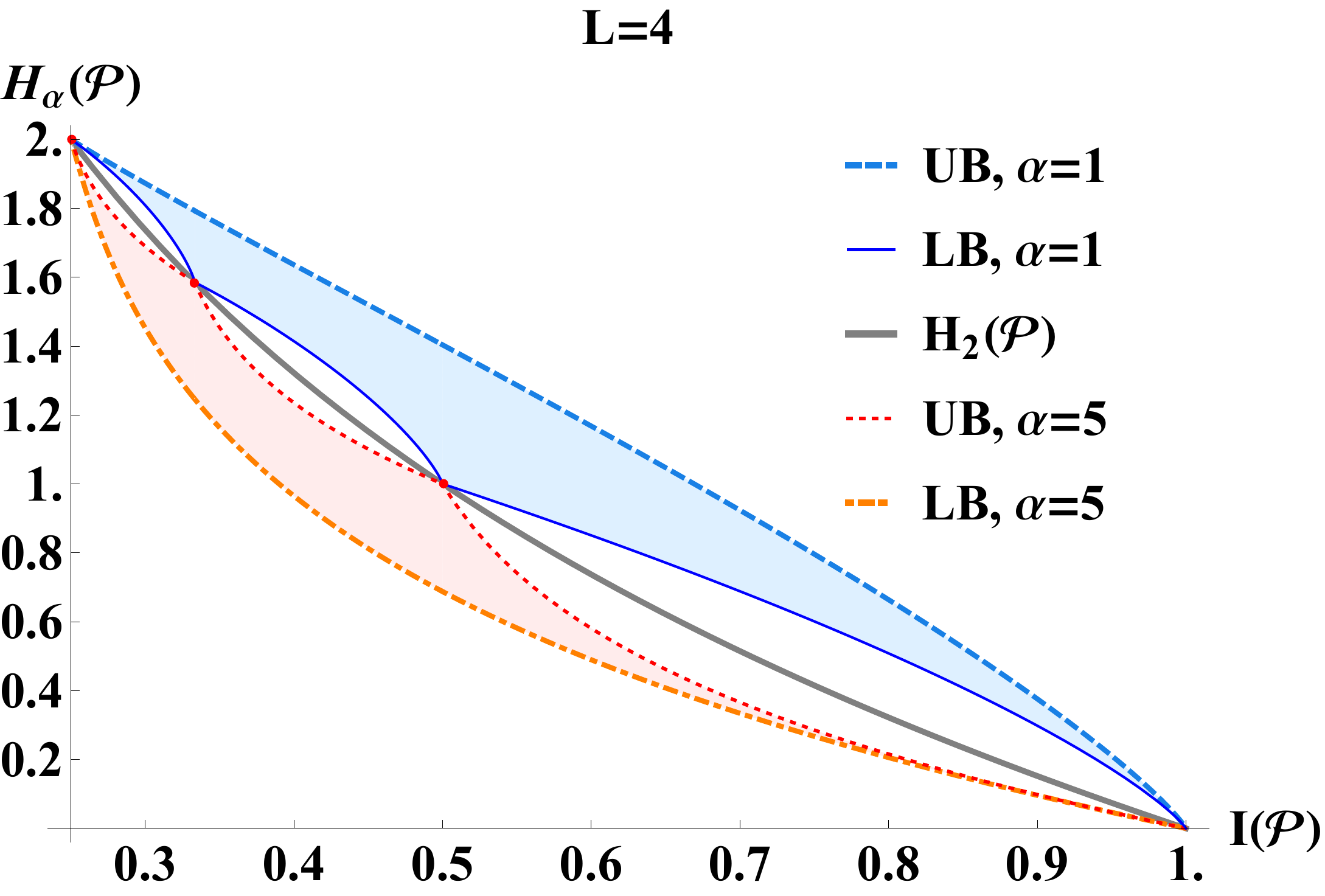}
\caption{Information diagrams of Shannon entropy and R\'{e}nyi 5-entropy (see also Refs. \cite{J3,HT}).} \label{rangeofentropy}
\end{figure}

\begin{figure}[H]
\centering\vspace{-0.4cm}
\includegraphics[width=0.48\textwidth]{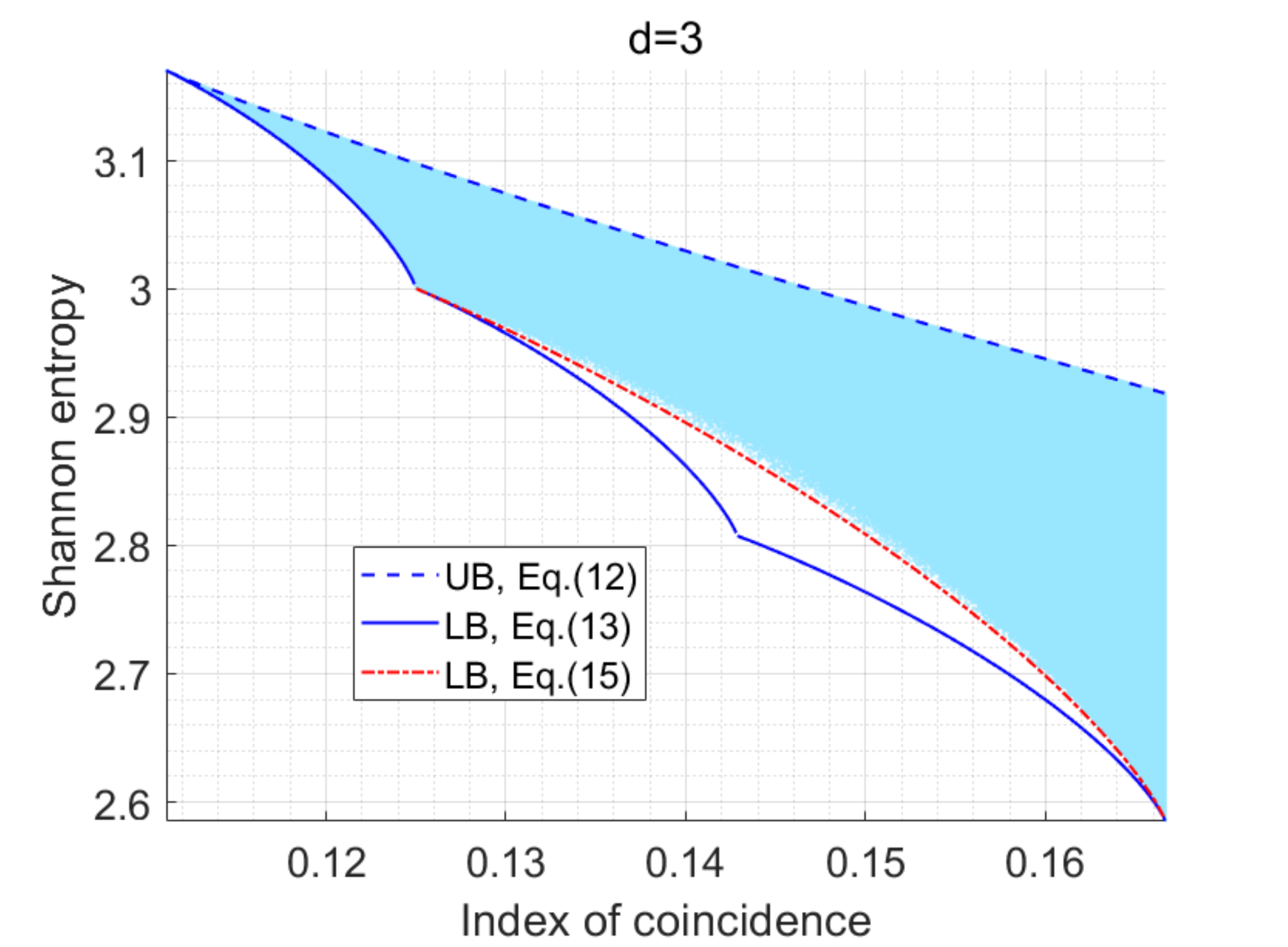}
\caption{Information diagram of Shannon entropy for SIC-POVM on $\mathcal{H}_3$ (cyan region).}\label{fig2_rangeforsicpovm}
\end{figure}
we obtain the information diagram shown in \myfig{fig2_rangeforsicpovm}.  It is not a surprise to see that our entropic lower bound for SIC-POVM is not tight when $Tr(\rho^2)>\frac{1}{2}$ since
\eqref{urforgsicpovm1} and \eqref{urforgsicpovm2} are based on \myref{indexgsic} only. Interestingly, the corresponding tight bound agrees with

\bgeq
H(\textbf{S}_g|\rho)\geq H\big(\mathscr{P}_y^4[2I(\textbf{S}_g|\rho)]\big)+1.
\edeq

\section{Uncertainty relations for MUMs}

\subsection{R\'{e}nyi entropy with $\alpha\leq 1$}

We show Theorem 2 in Appendix C.

\emph{Theorem 2.}
The sum of Shannon entropies for a finite set  {\bf P} of $|{\bf P}|$ MUMs with efficiency parameter $\kappa$ and performed on an arbitrary $d$-dimensional system $\rho$ is bounded from below by
\bgeq
H(\mathscr{P}_y^d[c])+k\log_2n+(|\textbf{P}|-k-1)\log_2(n+1)\label{Hmin},
\edeq
with $C(\textbf{P}|\rho)=\frac{|\textbf{P}|}{d}+\frac{\kappa d-1}{d(d-1)}\left[dTr(\rho^2)-1\right]$, here $n=\lfloor\frac{|\textbf{P}|}{C(\textbf{P}|\rho)}\rfloor$, $k=\lfloor (C(\textbf{P}|\rho)-\frac{M}{n+1})(n+1)n\rfloor$, and $c=C(\textbf{P}|\rho)-\frac{k}{n}-\frac{|\textbf{P}|-k-1}{n+1}$.

Despite the complex expression, this theorem can be understood in a simple way as is discussed in Appendix C. When $Tr(\rho^2)\in[\frac{1}{d},\frac{d+\kappa-2}{(d-1)^2}]$, (\ref{Hmin}) reduces to
\begin{equation}
(|\textbf{P}|-1)\log_2d+H\big(\mathscr{P}_y^d\big[C(\textbf{P}|\rho)-(|\textbf{P}|-1)/d\big]\big),
\label{tmumbound}
\end{equation}
which is actually valid for arbitrary R\'{e}nyi $\alpha$-entropy with $0<\alpha\leq1$, and quite similar to \eqref{urforgsicpovm1} it is tight.

We can linearize the first term of Eq.(\ref{Hmin}) based on its concavity with respect to $c$ as follows:\\
 $H(\mathscr{P}_y^d[c])\geq H(\mathscr{P}_y^d[\frac{1}{n+1}])$\\ $+n(n+1)(c-\frac{1}{n+1})\left[H(\mathscr{P}_y^d[\frac{1}{n}])-H(\mathscr{P}_y^d[\frac{1}{n+1}])\right]$,
which would then reduce to the result of Wu \emph{et al.} \cite{WYM} for MUBs,
\begin{align}
H({\bf B}|\rho)\geq&\left[|\textbf{B}|-nC(\textbf{B}|\rho)\right](n+1)\log_2(n+1)\nonumber\\
-&[|\textbf{B}|-(n+1)C(\textbf{B}|\rho)]n\log_2n\label{wym},
\end{align}

\begin{figure}[H]
\centering\vspace{-0.43cm}
\includegraphics[width=0.48\textwidth]{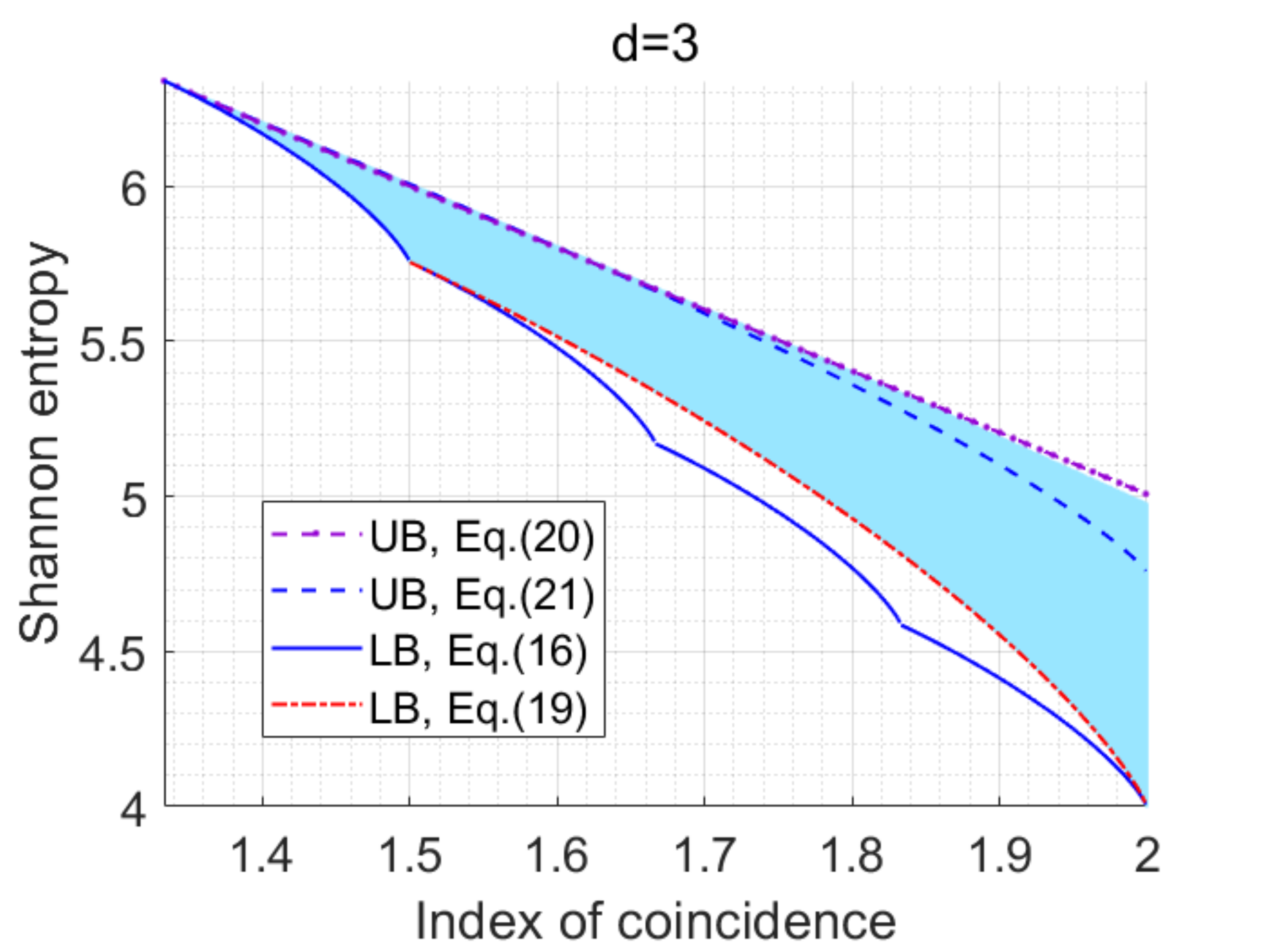}
\caption{Information diagram of Shannon entropy for complete MUBs in $\mathcal{H}_3$.}\label{fig3_rangeformub}
\end{figure}
where $C(\textbf{B}|\rho)$ is given by the right hand side of (\ref{indexmub}) and $n=\lfloor\frac{|\textbf{B}|}{C(\textbf{B}|\rho)}\rfloor$. (\ref{Hmin}) is generally improved from (\ref{wym}) and they are equivalent only when $c=\frac{1}{n}$ or $\frac{1}{n+1}$.

As can be seen in \myfig{fig3_rangeformub}, similar to \eqref{urforgsicpovm2}, \eqref{Hmin} is not tight when $Tr(\rho^2)>\frac{1}{2}$ for $d=3$ and the tight lower bound seems to be
\bgeq
H(\textbf{B}|\rho)\geq1+3H\big(\mathscr{P}_y^3[\frac{1+Tr(\rho^2)}{3}]\big).
\edeq
As for the upper bound, with $L=d$ we only propose the following two unproved approximations
\begin{gather}
(d+1)H\big(\mathscr{P}_x^d[I(\textbf{B}|\rho)/(d+1)]\big),\hspace{0.4em}\text{$Tr(\rho^2)\approx1$}\label{approx1}\\
d\log_2d+H(\mathscr{P}_x^d\left[I(\textbf{B}|\rho)-1\right]),\hspace{0.6em}Tr(\rho^2)\approx\frac{1}{d}\label{approx2}
\end{gather}

\subsection{R\'{e}nyi entropy with $\alpha\geq2$}

\emph{Theorem 3.}
Let \textbf{P} be a set of mutually unbiased measurements performed on a $d$-dimensional system $\rho$, then for any $\alpha\geq2$
\begin{align}
&\frac{1}{|\textbf{P}|}H_\alpha({\bf P}|\rho)\geq\frac{\alpha}{1-\alpha}\log_2p_{a}+\nonumber\\
& \frac{\log_2d}{(1-\alpha)\ln[1+(d-1)^{\frac{2}{\alpha}}]}\ln\left[1+(d-1)^{\frac{2}{\alpha}}\frac{p_{b}^2}{p_{a}^2}\right],
\label{alpha2}
\end{align}
where $p_a=\frac{1+\sqrt{(d-1)(dc-1)}}{d}$, $p_b=\frac{1-\sqrt{(dc-1)/(d-1)}}{d}$, and with $C(\textbf{P}|\rho)$ being the right hand side of \myref{indexmum}, $c=\frac{1}{|\textbf{P}|}C(\textbf{P}|\rho)$.

This inequality is a direct result of the fact that the right hand side of (\ref{alpha2}) is convex with respect to $c$. When $\alpha>2$, \myref{alpha2} is improved from Rastegin's lower bounds $L_{Ras1}$ \cite{Raste} and $L_{Ras2}$ \cite{design2}
\bgeq\left\{\begin{aligned}
&L_{Ras1}=\frac{\alpha}{2(1-\alpha)}\log_2c\\
&L_{Ras2}=\frac{\alpha-2}{1-\alpha}\log_2\Big(\frac{1+\sqrt{(dc-1)(d-1)}}{d}\Big)\\
+&\frac{1}{1-\alpha}\log_2c,
\end{aligned}\right.
\edeq
\begin{figure}[H]
\centering\vspace{0.2cm}
\includegraphics[width=0.48\textwidth]{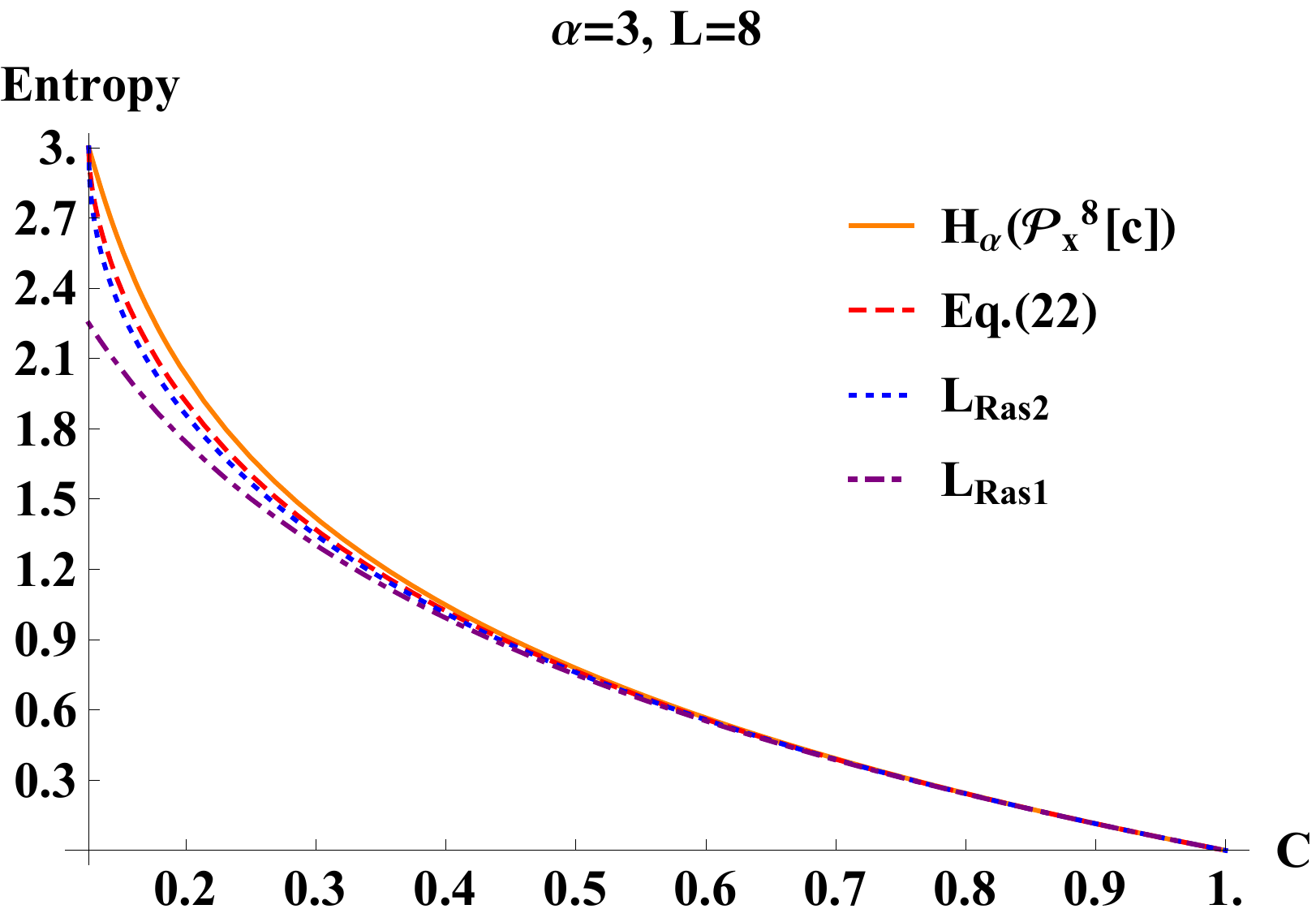}
\caption{Lower bound on R\'{e}nyi-3 entropy of a single probability distribution.}
\label{lowerboundcurves}
\end{figure}
and when $\alpha=2$ they all reduce to $-\log_2c$. A comparison between these results when $\alpha=3$ and $L=8$ is shown in \myfig{lowerboundcurves}

\subsection{Entropy region}

The entropies of performing a finite ordered set of generalized measurements $\textbf{P}=\{\mathcal{P}^m\}$ on a $d$-dimensional system described by $\rho$ form an vector, the $m{th}$ element of which is  $H(\mathcal{P}^m|\rho)$. The region of all possible entropic vectors induced by \textbf{P} is called the entropy region of \textbf{P}. The entropy region of a given measurement set contains much more information besides the entropic lower bound, and we expect it to be as meaningful in quantum information theory as in the classical counterpart.

\begin{figure}[ht]
\centering
\subfigure[Shannon entropy region for three MUBs.]{
\begin{minipage}[h]{0.45\linewidth}
\centering
\includegraphics[width=1\linewidth,height=10cm]{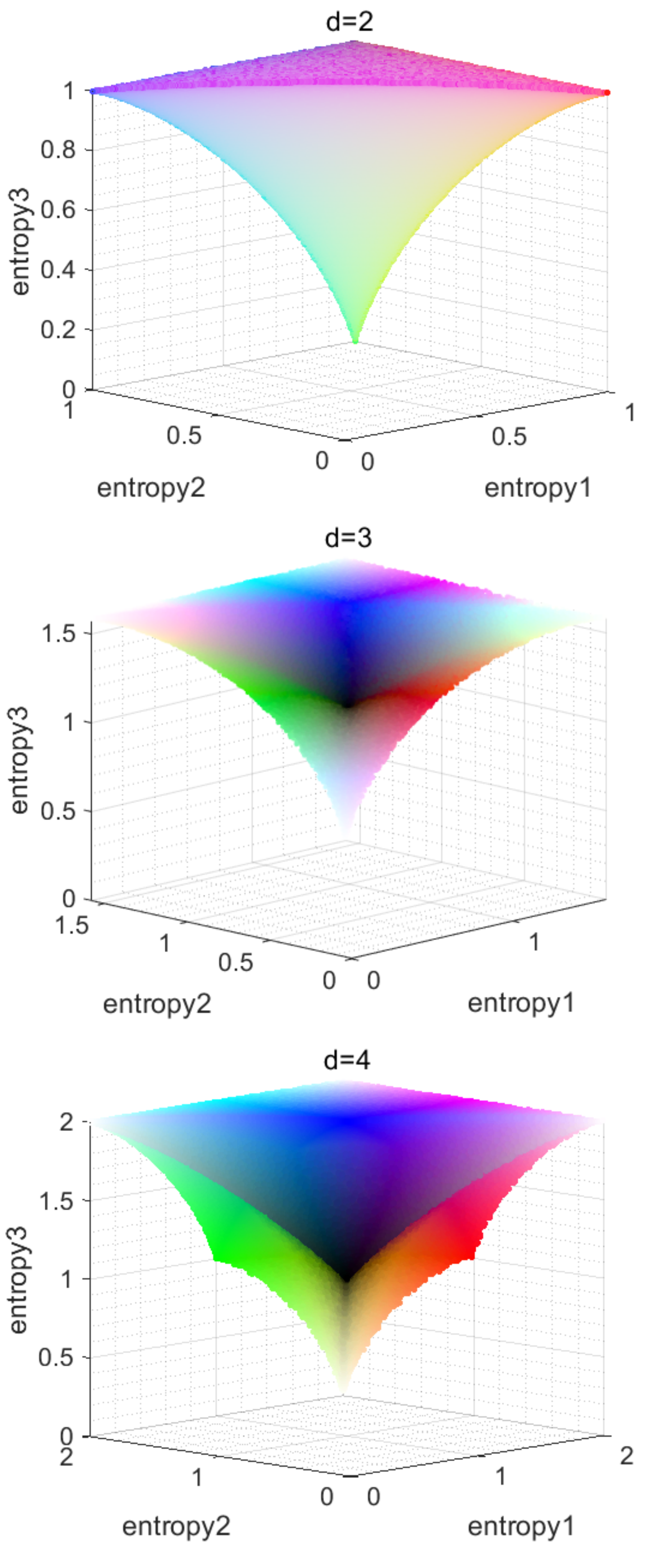}
\end{minipage}%
}%
\quad
\subfigure[Shannon entropy region of three probability distributions satisfying \myref{index3mub}]{
\begin{minipage}[h]{0.45\linewidth}
\centering
\includegraphics[width=1\linewidth,height=10cm]{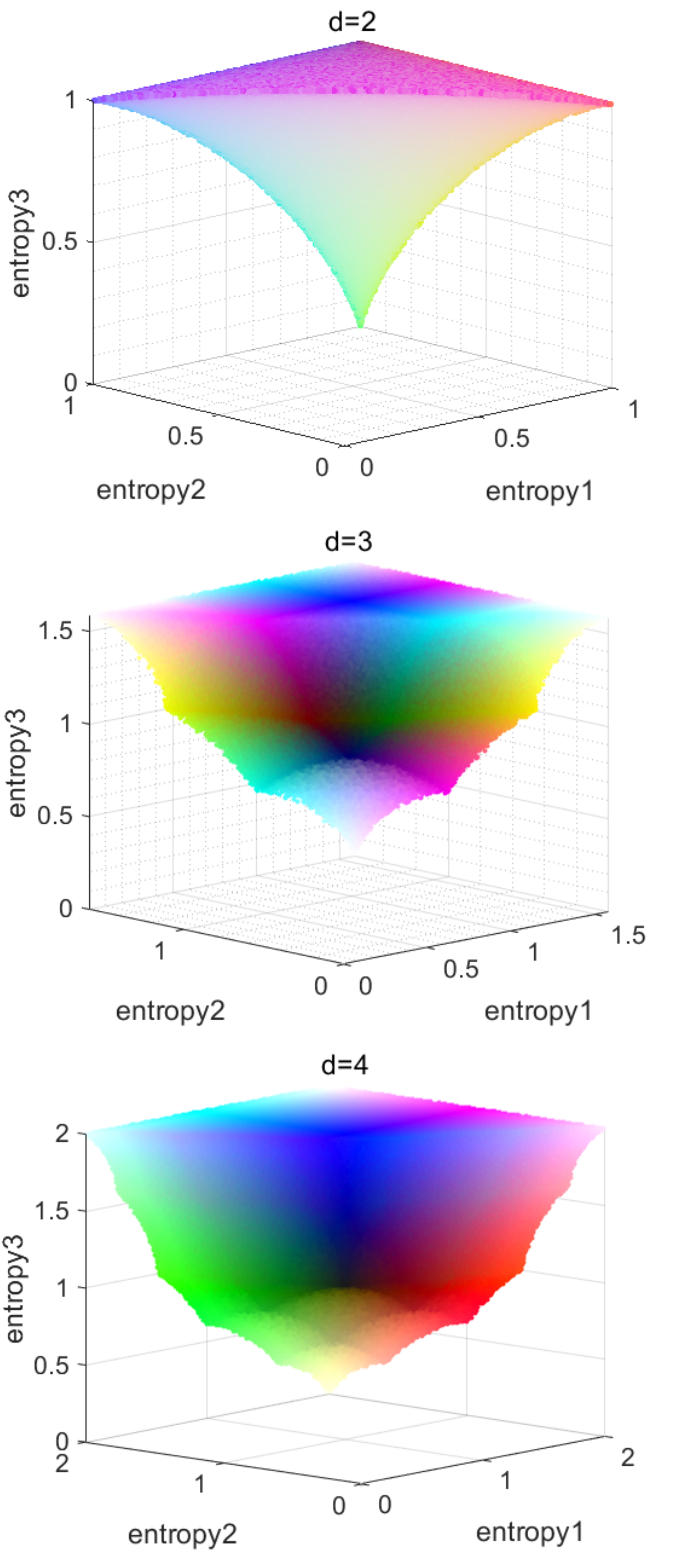}
\end{minipage}
}%
\centering
\caption{Shannon entropy regions for three MUBs}\label{entropyregion}
\end{figure}

We make a comparison here between the Shannon entropy region for three MUBs in $\mathcal{H}_d$ and that of three probability distributions over $d$ outcomes satisfying
\bgeq
\frac{3}{d}\leq\sum_{m=1}^3I(\mathscr{P}^m)\leq\max\limits_{\rho}\{I(\textbf{B}|\rho)\}=1+\frac{2}{d}\label{index3mub}
\edeq

\begin{figure}[H]
\centering\vspace{-0.cm}
\includegraphics[width=0.48\textwidth]{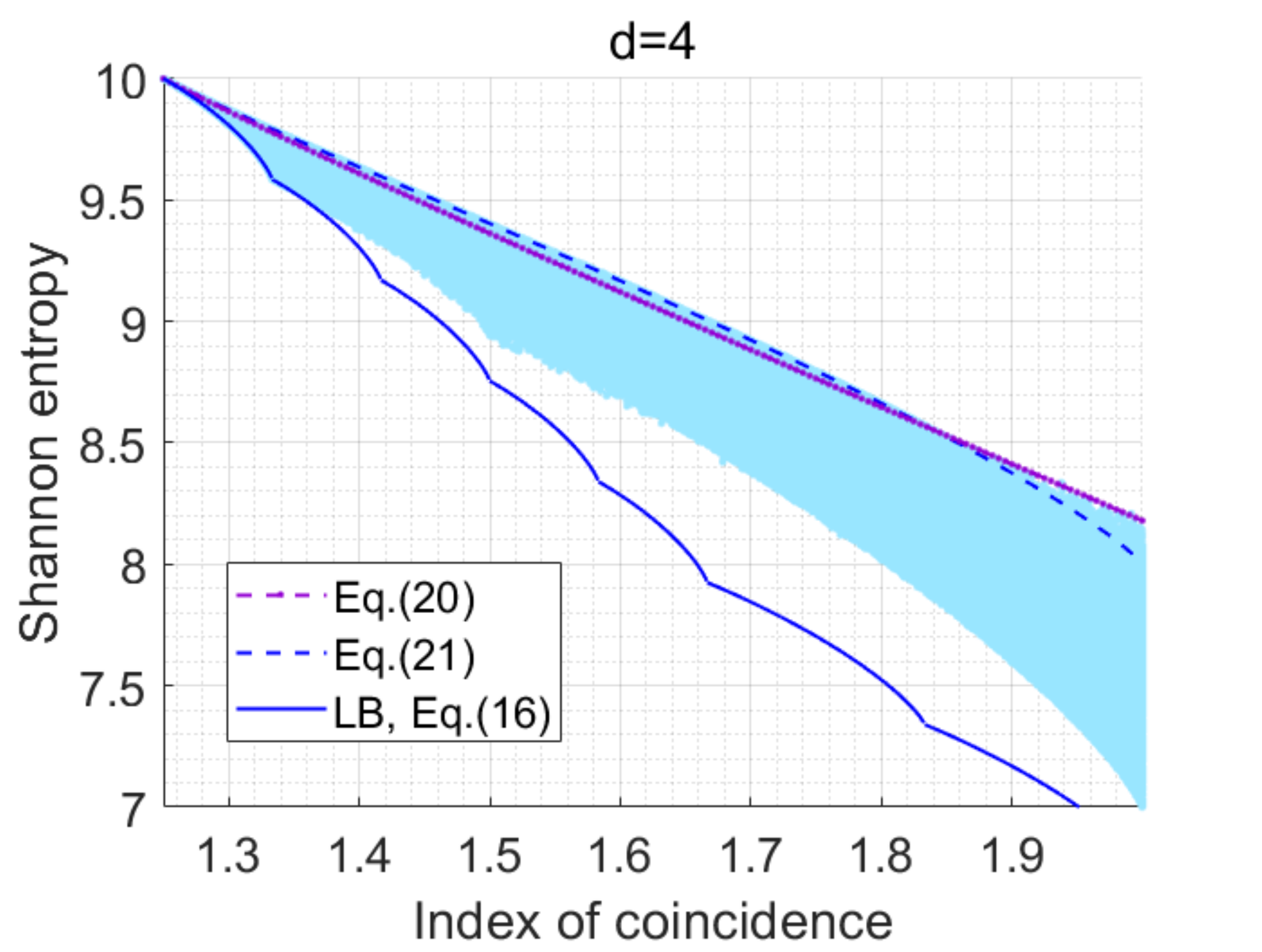}
\caption{Information diagram of Shannon entropy for complete MUBs in $\mathcal{H}_4$.}\label{fig6_rangeformub}
\end{figure}
As can be seen in Fig.\hspace{0.3em}\ref{entropyregion}, the entropy region of probability distributions satisfying \myref{index3mub} is the same to that for three MUBs when $d=2$, while in higher dimensions distinctions show up at places where the sum of entropies is relatively small, which is in accordance with the information diagrams.

\section{Discussions}

We can see from Figs.\hspace{0.3em}(\ref{fig2_rangeforsicpovm},\ref{fig3_rangeformub}, and \ref{fig6_rangeformub}) that the tight lower R\'{e}nyi entropic ($\alpha<2$) bound curves for both complete MUBs and SIC-POVMs are non-differentiable
at $Tr(\rho^2)=\frac{1}{k}$  $(\forall k=2,\cdots,d-1)$, which divide the curves into $d-1$ sections. A natural thought is that different sections corresponds with density matrices at different boundaries of the set of positive semi-definite Hermitian matrices, namely, different sections of the lower bound curve are attained by density matrices of different ranks.

{\bf Conjecture}. The tight lower bound on Shannon entropy for complete MUBs or SIC-POVMs on $\mathcal{H}_d$ can only be achieved by density matrices satisfying $(\lambda_1,\lambda_2,\cdots)=\mathscr{P}_y^d[Tr(\rho^2)]$, where $\{\lambda_i\}$ are nonzero eigenvalues of $\rho$ and arranged in descending order.

We believe this conjecture, if confirmed, will be helpful in searching for tight state-independent EURs for complete MUBs and SIC-POVMs, which could be more efficient in applications of quantum information theory.

Based on the conjecture above, we have an alternative form of \myref{Hmin} for MUBs when $Tr(\rho^2)\leq\frac{1}{d-1}$

\begin{align}
H({\bf B}|\rho)\geq (|\textbf{B}|-1)\log_2d-Tr[\rho\log_2\rho],
\end{align}
which coincides with the uncertainty relation for two observables proposed by Berta \emph{et al.} \cite{berta}.

Lastly, we show an application of entropic uncertainty relations in entanglement detection. Let $\rho_{AB}=\sum_ip_i\rho_i^A\otimes\rho_i^B$ be an arbitrary separable state on the bipartite Hilbert space $\mathcal{H}_{A}\otimes\mathcal{H}_{B}$, and $\rho_A=Tr_B(\rho_{AB})$ and $\rho_B=Tr_A(\rho_{AB})$ are the reduced density matrices. According to the results shown in \mycite{entangle2}, for any nondegenerate observables $\{A_1,A_2,\cdots\}$ on $\mathcal{H}_{A}$ and $\{B_1,B_2,\cdots\}$ on $\mathcal{H}_{B}$, there is
\begin{align}
\sum_mH_\alpha&\big(\mathscr{P}(A_m\otimes B_m|\rho_{AB})\big)\geq D=\max\{D_A,D_B\},\nonumber\\
\text{where}\hspace{0.5em}&D_A=\sum_mH_\alpha\big(\mathscr{P}(A_m|\rho_A)\Big),\label{separ}\\
&D_B=\sum_mH_\alpha\big(\mathscr{P}(B_m|\rho_B)\big)\nonumber. \hspace{0.5em}(\alpha>0)
\end{align}
Here, $\mathscr{P}(A_m\otimes B_m|\rho_{AB})$ is the probability distribution induced by measuring $A_m\otimes B_m$ on $\rho_{AB}$, and $\mathscr{P}(A_m|\rho_A)$ and $\mathscr{P}(B_m|\rho_B)$ are the two corresponding marginal probability distributions.

Any bipartite state violating \myref{separ} must be entangled. When $\{A_m\}$ and $\{B_m\}$ are complementary observables, $D$ is the sum of entropies for measuring $\rho_A$ or $\rho_B$ in MUBs. According to our uncertainty relations in the previous section, a further lower bound for $D$ can be obtained directly from \eqref{alpha2} when $\alpha\geq2$ and from \eqref{Hmin} when $\alpha=1$.

As an example, consider a pair of qudits in a Werner state $\mathcal{W}_{AB}=\frac{1-p}{d^2}{\bf 1}_{d^2}+p|\psi\rangle\langle\psi|$ $(0\leq p\leq1)$, where $|\psi\rangle=\frac{1}{\sqrt{d}}(|0\rangle\otimes|0\rangle+\cdots+|d-1\rangle\otimes|d-1\rangle)$, and $\{\sigma_m\}$ ($m=1,2,\cdots,M)$ is a set of complementary observables in $\mathcal{H}_d$. The density matrix of a single qudit is then $\mathcal{W}_A=\mathcal{W}_B=\frac{1}{d}{\bf 1}_d$, which is independent of $p$. For simplicity, suppose now $d$ is a prime power and $M=d+1$, in which case $D=(d+1)\log_2d$ is state-independent, and moreover, the strongest form of \eqref{separ} becomes
\bgeq
\sum_m^{d+1}H_\infty\big(\mathscr{P}(\sigma_m\otimes\sigma_m|\mathcal{W}_{AB})\big)\geq(d+1)\log_2d\label{separa}
\edeq
since R\'{e}nyi $\alpha$-entropy is a non-increasing function of $\alpha$. Numerical results show that \eqref{separa} can be violated for $p>0.33$ when $d=2$ and for $p>0.46$ when $d=3$. As the bipartite Werner state is entangled if and only if $p>\frac{1}{d+1}$ \cite{sep2,sep3,sep4,sep5}, \myref{separa} is strong enough when $d=2$ but it is not strong when $d=3$ and fails to detect all entangled states.

From the above example we know that our EURs can be used to detect entanglement, and more stronger separability criteria based on our uncertainty relations are also possible. More works on entropic separability criteria can be found in Refs. [11, 15, 23].

\section*{CONCLUSION}
In this paper we have obtained improved entropic uncertainty relations for general symmetric informationally complete positive operator-valued measures and mutually unbiased measurements in terms of R\'{e}nyi entropy, which are shown to be tight for sufficiently mixed states. It might be the first time that tight state-dependent entropic uncertainty relations for multiple generalized measurements have been obtained. By random sampling density matrices and calculating the corresponding entropy for a given set of measurements, comparisons between our entropic bounds and the numerical optimal bounds are made via information diagrams. Our investigation of entropic uncertainty relations could provide some insights for further applications of uncertainty relations in information theory.

\acknowledgments

This work is supported by the National Key R\&D Program of China (Grants No. 2017YFA0303703 and No. 2016YFA0301801)
and the National Natural Science Foundation of China (Grant No. 11475084).

\appendix

\begin{widetext}

\section{Proof of Theorem 1}
We employ Lagrangian multiplier method to find the distributions which make R\'{e}nyi $\alpha$-entropy attain local extreme values on the set: ${\bf S}(L,c)=\left\{\mathscr{P}|length(\mathscr{P})=L\geq3,\hspace{0.5em}I(\mathscr{P})=c<1\right\}$.
\begin{align}
L_\alpha&=\sum_{i=1}^L{p_{i}}^\alpha+\lambda(\sum_{i=1}^Lp_{i}^2-c) +\lambda'(\sum_{i=1}^Lp_{i}-1);\nonumber\\
\frac{\partial{L_\alpha}}{\partial{p_{i}}}&=\alpha p_i^{\alpha-1}+2\lambda p_{i}+\lambda'=0, \hspace{0.5em}1\leq i\leq L.\hspace{1em}(\alpha>0,\alpha\neq1)\label{lagrange}
\end{align}
where $\lambda$ and $\lambda'$ are multipliers. Note the equation $\alpha x^{\alpha-1}=-2\lambda x-\lambda'x$ has at most two different solutions as $x^{\alpha-1}$ is either concave or convex with respect to $x$ and $-2\lambda x-\lambda'x$ describes a line. But \eqref{lagrange} is not valid if there exists $1\leq i\leq L$ such that $p_i=0$.

When $\alpha<1$ and $p_i\to0$, in which case $p_i^{\alpha-1}\to+\infty$. This implies on ${\bf S}(L,c)$, R\'{e}nyi $\alpha$-entropy can only attain local minimum value at probability distributions whose positive probabilities satisfy \eqref{lagrange}, and it can never attain local maximum value at a distribution the smallest probability in which is 0. When $\alpha>1$, $\frac{\partial}{\partial p_i}(\sum_ip_i^2-c)\big |_{p_i=0}=0$, again, \eqref{lagrange} is a restriction on the positive probabilities only.

We only need to consider those distributions containing at most two different positive probabilities, $p_a$ and $p_b$ say, and let's parameterize  them with three parameters as follows: N, the number of positive probabilities; $N_a$, the number of probabilities being $p_a$; $c$, the index of coincidence. We arrange the positive probabilities in descending order and represent the distribution formally as
\begin{equation}
\mathscr{P}[c,N,N_a]=\Big(N_a\odot p_a,(N-N_a)\odot p_b\Big),\hspace{1em}(1/N\leq c\leq1/N_a)\label{edistribution}
\end{equation}
where $N_a\odot p_a$ is shorthand for $N_a$ probabilities being $p_a$. Combined with the condition that $N_a p_a+(N-N_a) p_b=1$ and $I\big( \mathscr{P}[c,N,N_a]\big)=c$, we have $p_a=$ $\frac{1+\sqrt{(Nc-1)(N-N_a)/N_a}}{N}$ and $p_b=\frac{1-\sqrt{(Nc-1)N_a/(N- N_a)}}{N}$. It can be checked that $\mathscr{P}[c_1,N,N_a]$ majorizes $\mathscr{P}[c_2,N,N_a]$ if $c_1>c_2$, thus $H_\alpha(\mathscr{P}[c,N,N_a])$ is a decreasing function of $c$.

Given the values of $u$, $c$ $(1/N\leq c<\frac{1}{N_a})$, and $N$ ($N_a$), the values of $N_a$ $(N)$, $p_a$, and $p_b$, if exist, are uniquely determined by \eqref{distri}.
\begin{gather}\left\{\begin{aligned}
0\leq p_b<p_a,N\geq N_a\geq1,0<\alpha\neq2\hspace{0.5em}\text{and}\hspace{0.5em}\alpha\neq 1,\hspace{5.5cm} \\
N_ap_a+(N-N_a)p_b=1;\hspace{1em}N_ap_a^2+(N-N_a)p_b^2=c;\hspace{1em}N_ap_a^\alpha+(N-N_a)p_b^\alpha=u>0.\\
\end{aligned}\right.\label{distri}\\
H_\alpha\big(\mathscr{P}[1/N+s,N,N_a]\big)=\log_2N-\frac{\alpha Ns}{2\ln2}+\frac{\alpha(\alpha-2)N^{\frac{3}{2}}
N_as^{\frac{3}{2}}}{4\ln2(N-N_a)}+o(s^2),\hspace{1em}(0<s\ll\frac{1}{N_a}-\frac{1}{N}).\label{expansion}
\end{gather}

Note that $H_\alpha\big(\mathscr{P}[\frac{1}{N},N,N_a]\big)=\log_2N$ is independent of $N_a$ and $\lim_{c\to1/N_a}H_\alpha\big(\mathscr{P}[c,N,N_a]\big)=\log_2N_a$ is independent of $N$, thus $H_\alpha\big(\mathscr{P}[c,N,N_a]\big)$ is monotonic of $N$ if $c\neq\frac{1}{N}$ and monotonic of $N_a$ if $c\neq \frac{1}{N_a}$, more concretely, taking the series expansion of entropy \eqref{expansion} into consideration we have
\begin{gather}
(2-\alpha)H_\alpha\big(\mathscr{P}[c,N,N_{a}]\big)\geq(2-\alpha)H_\alpha\big(\mathscr{P}[c,N,N_{a}']\big),
\hspace{0.5em}(N_{a}< N_{a}', \frac{1}{N}\leq c\leq \frac{1}{N_{a}'});\label{1}\\
(2-\alpha)H_\alpha\big(\mathscr{P}[c,N,N_a]\big)\leq(2-\alpha)H_\alpha\big(\mathscr{P}[c,N',N_a]\big),
\hspace{0.5em}(N< N',\frac{1}{N}\leq c\leq\frac{1}{N_a}).\label{2}
\end{gather}
We can conclude from \eqref{1} and \eqref{2} that for any distribution $\mathscr{P}$ over $L$ outcomes with $I(\mathscr{P})=c$ there is
\bgeq
(2-\alpha)H_\alpha\big(\mathscr{P}[c,L,1]\big)\geq(2-\alpha)H_\alpha\big(\mathscr{P}\big)\geq(2-\alpha) H_\alpha\big(\mathscr{P}[c,N,N-1]\big),\label{optprodis}
\edeq
where $N$ is an integer such that $\frac{1}{L}\leq\frac{1}{N}\leq c<\frac{1}{N-1}$, namely, $N=\lceil\frac{1}{c}\rceil$. This completes the proof of Theorem 1.

\section{Properties of extreme values}

Let's reparametrize $\mathscr{P}[c,N,N_a]$ as $\mathscr{P}^*[c,N,\theta]$, where $\theta=2\arccos\sqrt{N_a/N}$ and $\theta\in[0,\pi)$. We have
\begin{equation*}\begin{aligned}
H_\alpha\big(\mathscr{P}^*[c,N,\theta]\big)&=\frac{1}{1-\alpha}\log_2
\Big[
N\cos^2\frac{\theta}{2}\Big(\frac{1+\sqrt{Nc-1}\tan\frac{\theta}{2}}{N}\Big)^\alpha+
N\sin^2\frac{\theta}{2}\Big(\frac{1-\sqrt{Nc-1}\cot\frac{\theta}{2}}{N}\Big)^\alpha
\Big]\\
&=\frac{1}{1-\alpha}\log_2M_\alpha(\mathscr{P}^*[c,N,\theta])\\
(\alpha-1)\frac{\partial^2 }{\partial c^2}H_\alpha\big(\mathscr{P}^*[c,N,\theta]\big)
&=f(\alpha,z,\theta)\frac{\alpha N^{\alpha+1}\sin\frac{\theta}{2}\cos^3\frac{\theta}{2}}{4\ln2(Nc-1)^{3/2}M_\alpha^2\big(\mathscr{P}^*[c,N,\theta]\big)}
\left(1+\sqrt{Nc-1}\tan\frac{\theta}{2}\right)^{2\alpha-2}
\end{aligned}\end{equation*}
where, with $z=\frac{1-\sqrt{Nc-1}\cot\frac{\theta}{2}}{1+\sqrt{Nc-1}\tan\frac{\theta}{2}}$ $(0<z\leq\tan\frac{\theta}{2})$,  $f(\alpha,z,\theta)$
$$\begin{aligned}
=&\frac{2\tan^2\frac{\theta}{2}}{1+z\tan^2\frac{\theta}{2}}(z^{\alpha-1}-1)^2
+z^{\alpha-1}\left[
-\tan^2\frac{\theta}{2}z^{\alpha-1}+z^{1-\alpha}+(\alpha-1)(z\tan^2\frac{\theta}{2}-\frac{1}{z})+(2-\alpha)(\tan^2\frac{\theta}{2}-1)
\right]\nonumber\\
\geq &z^{\alpha-1}\left[-\tan^2\frac{\theta}{2}z^{\alpha-1}+z^{1-\alpha}
+(\alpha-1)(z\tan^2\frac{\theta}{2}-\frac{1}{z})+(2-\alpha)(\tan^2\frac{\theta}{2}-1)\right]
\end{aligned}$$
when $\alpha\in(0,1)\cup[2,+\infty)$ and $0<\tan\frac{\theta}{2}\leq1$,
\bgeq
f(\alpha,z,\theta)\geq z^{\alpha-1}\left[-z^{\alpha-1}+z^{1-\alpha} +(\alpha-1)(z-\frac{1}{z})\right]\geq0\Longrightarrow
(\alpha-1)\frac{\partial^2 }{\partial c^2}H_\alpha\big(\mathscr{P}^*[c,N,\theta]\big)\geq0\label{convexity1}
\edeq
As for Shannon entropy, when $0<z<\tan\frac{\theta}{2}\leq1$
\begin{align}
&\frac{\partial^2}{\partial c^2}H\big(\mathscr{P}^*[c,N,\theta]\big)=\log_2(1+z\tan{\frac{\theta}{2}})-\log_2(1-z\cot{\frac{\theta}{2}})
-\frac{1}{1-z\cot{\frac{\theta}{2}}}+\frac{1}{1+z\tan{\frac{\theta}{2}}}\leq0.\label{convexity2}
\end{align}

\eqref{convexity1} and \eqref{convexity2} imply that when $N_a\in [N/2,N]$, $H_\alpha\big(\mathscr{P}[c,N,N_a]\big)$ is concave with respect to $c$ when $\alpha\leq1$ and convex with respect to c when $\alpha\geq2$.

\section{Proof of Theorem 2}
Let {\bf g}=$\{\mathscr{P}^g\}$ denote the probability distributions at which $\sum_{m=1}^MH(\mathscr{P}^m)$ is minimum under the restriction
\bgeq
\forall 1\leq m\leq M,length(\mathscr{P}^m)=L;\hspace{0.5em}\sum_{m=1}^{M}I(\mathscr{P}^m)=\sum_mc^m=c\hspace{0.5em}\text{($c$ is a constant, $c\in[\frac{M}{d},M]$)},\label{restriction}
\edeq
where $\mathscr{P}^g$ is the $gth$ distribution in {\bf g}. Firstly, according to \eqref{optprodis} (or Theorem 1) and \eqref{convexity2} we have the following \\
{\bf Property 1}. $\mathscr{P}^g$ must be in the form $\mathscr{P}^g=\mathscr{P}_y^L[c^g]$ for any $g$.\\
{\bf Property 2}. At most one element in \textbf{g}, $\mathscr{P}^k$ say, is not uniform in its nonzero part.\\
It can be proved that for any $2\leq m<n$ $(n,m\in N^+)$,
\begin{gather}\left\{\begin{aligned}
1.\hspace{0.5em}H\big(\mathscr{P}_y^L[1/n]\big)+H\big(\mathscr{P}_y^L[1/m+s]\big)>H\big(\mathscr{P}_y^L[1/m]\big)+
H\big(\mathscr{P}_y^L[1/n+s]\big),\hspace{0.5em} 0\leq s\leq1/n/(n-1)\\
2.\hspace{0.5em}H\big(\mathscr{P}_y^L[1/n-s]\big)+H\big(\mathscr{P}_y^L[1/m]\big)\geq
H\big(\mathscr{P}_y^L[1/m-s]\big)+H\big(\mathscr{P}_y^L[1/n]\big),\hspace{0.5em}0\leq s\leq1/n/(n+1)
\end{aligned}\right.\label{last}\\
\text{note here:}\hspace{0.5em} IC(\mathscr{P}_y^L[1/n])+IC(\mathscr{P}_y^L[1/m+s])=IC(\mathscr{P}_y^L[1/n+s])+IC(\mathscr{P}_y^L[1/m])\nonumber\\
\hspace{5em}IC(\mathscr{P}_y^L[1/n-s])+IC(\mathscr{P}_y^L[1/m])=IC(\mathscr{P}_y^L[1/m-s])+IC(\mathscr{P}_y^L[1/n]),\nonumber
\end{gather}
with $N_g$ denoting the number of nonzero probabilities of $\mathscr{P}^g$, a direct result of {properties 1-2} and \eqref{last} is the following\\
$\begin{aligned}
\text{{\bf Property 3}}.\hspace{1em}1.\hspace{0.5em}&\max\limits_{g,g'}|\{N_g-N_{g'}\}|\leq1;\hspace{2em}
2.\hspace{0.5em}\text{if}\hspace{0.5em} N_k=\min\limits_{g}\{N_g\}, \hspace{0.5em}\text{then}\hspace{0.5em} \forall g,N_g-N_k=0
\end{aligned}$\\
With { Properties 1-3}, it's enough to determine {\bf g} (Theorem 2). To show the first inequality of \eqref{last} we only need to show $\log_2N-H\big(\mathscr{P}_y[1/N+s,N,N-1]\big)$ is an increasing function of N. Under the parametrization introduced in Appendix B we have,
\begin{align}
&\log_2N-H\big(\mathscr{P}^*[1/N+s,N,\theta]\big)
=\cos^2\frac{\theta}{2}\Big(1+\sqrt{Ns}\tan\frac{\theta}{2}\Big)
\log_2\Big[\cos^2\frac{\theta}{2}\Big(1+\sqrt{Ns}\tan\frac{\theta}{2}\Big)\Big]\\
&\sin^2\frac{\theta}{2}\Big(1-\sqrt{Ns}\cot\frac{\theta}{2}\Big)
\log_2\Big[\sin^2\frac{\theta}{2}\Big(1-\sqrt{Ns}\cot\frac{\theta}{2}\Big)\Big]=h\big(s,N,\theta\big)\nonumber
\end{align}
Let $\theta_y(N)=2\arctan\frac{1}{\sqrt{N-1}}$, then $\mathscr{P}[c,N,N-1]=\mathscr{P}^*[c,N,\theta_y(N)]$,
\begin{align}
&\frac{\partial}{\partial N}h\big(s,N,\theta\big)=\frac{s}{N}\frac{\partial}{\partial s}h\big(s,N,\theta\big)>0,\hspace{0.5em}\frac{\partial}{\partial \theta}h\big(s,N,\theta\big)<0,\hspace{0.5em}\frac{d \theta_y}{d N}<0\nonumber\\
&\frac{\partial}{\partial N}h\big(s,N,\theta_y(N)\big)
=\Big[\frac{\partial}{\partial N}h\big(s,N,\theta\big)+
\frac{\partial}{\partial \theta}h\big(s,N,\theta\big)\frac{d\theta_y}{d N}\Big]\Big|_{\theta=\theta_y(N)}\geq0
\label{firstineq}
\end{align}
Hence $h(s,N,\theta_y(N))$ is an increasing function of N, and the second inequality of \eqref{last} can be proved similarly.

It turns out that {\bf g} is also the set of probability distributions that descends entropy the fastest locally.
Consider $c=\frac{M}{L}$ (this is when probability distributions are all uniform) in the beginning and then let $c$ increase, then according to { Properties 1, 2} and \eqref{expansion} obviously the steepest descent of Shannon entropy is given by

$$
\left.\begin{aligned}
&\Big\{(M-1)\odot\mathscr{P}_y^L\big[\frac{1}{L}\big],\mathscr{P}_y^L\big[c-\frac{M-1}{L}\big]\Big\},
\hspace{9em}\frac{M}{L}\leq c\leq \frac{M-1}{L}+\frac{1}{L-1}\\
&\Big\{(M-2)\odot\mathscr{P}_y^L\big[\frac{1}{L}\big],\mathscr{P}_y^L\big[\frac{1}{L-1}\big],\mathscr{P}_y^L\big[c-\frac{M-2}{L}-\frac{1}{L-1}\big]\Big\},
\frac{M-1}{L}+\frac{1}{L-1}\leq c\leq \frac{M-2}{L}+\frac{2}{L-1}\\
&\hspace{4em}\cdots\cdots\\
&\Big\{(M-1)\odot\mathscr{P}_y^L\big[\frac{1}{L-1}\big],\mathscr{P}_y^L\big[c-\frac{M-1}{L-1}\big]\Big\},
\hspace{7.5em}\frac{M}{L-1}\leq c\leq \frac{M-1}{L-1}+\frac{1}{L-2}\\
&\hspace{4em}\cdots\cdots
\end{aligned}\right\}={\bf g}
$$
where $M\odot \mathscr{P}$ is shorthand for $M$ probability distributions being $\mathscr{P}$.
\end{widetext}

\end{document}